\documentclass[preprint,aoas]{imsart}

\RequirePackage[OT1]{fontenc}
\RequirePackage{amsthm,amsmath}
\RequirePackage{natbib}
\RequirePackage{graphicx}
\RequirePackage[colorlinks,citecolor=blue,urlcolor=blue]{hyperref}
\RequirePackage{hypernat}

\arxiv{math.PR/0000000}
\startlocaldefs
\numberwithin{equation}{section}
\theoremstyle{plain}

\endlocaldefs

\usepackage[english]{babel}
\usepackage[latin1]{inputenc}
\numberwithin{equation}{section}
\usepackage[hang,small,tight]{subfigure}
\usepackage{amssymb,amsmath,natbib}
\usepackage{graphicx}
\usepackage{lscape}
\usepackage{fullpage}
\usepackage{bm}
\usepackage{multirow}
\usepackage{xspace,colortbl}

\newcommand{\bfalpha}{\mbox{\boldmath $\alpha$}}
\newcommand{\bfSigma}{\mbox{\boldmath $\Sigma$}}

\newcommand{\bftheta}{\mbox{\boldmath $\theta$}}

\newcommand{\bfbeta}{\mbox{\boldmath $\beta$}}
\newcommand{\bfeta}{\mbox{\boldmath $\eta$}}

\newcommand{\bflambda}{\mbox{\boldmath $\lambda$}}

\newcommand{\bfgamma}{\mbox{\boldmath $\gamma$}}

\begin{document}

\begin{frontmatter}
\title{Population counts along elliptical habitat contours: hierarchical modelling using Poisson-lognormal mixtures with nonstationary spatial structure}
\runtitle{Nonstationarity in population counts models}

\begin{aug}
\author{\fnms{} \snm{Alexandra M. Schmidt}\thanksref{m1}\ead[label=e1]{alex@im.ufrj.br}},
\author{\fnms{} \snm{Marco A. Rodr\'iguez}\thanksref{m2}\ead[label=e2]{marco.rodriguez@uqtr.ca}}
\and
\author{\fnms{} \snm{Estelina S. Capistrano}\thanksref{m1}
\ead[label=e3]{estelina@dme.ufrj.br}
\ead[label=u1,url]{http://www.dme.ufrj.br/$\sim$alex}
\ead[label=u2,url]{http://www.uqtr.ca/marco.rodriguez}}

\runauthor{Schmidt, Rodr\'iguez and Capistrano}

\affiliation{Universidade Federal do Rio de Janeiro, Brazil,\thanksmark{m1} \\ Universit\'e du Qu\'ebec \`a Trois-Rivi\`eres, Canada\thanksmark{m2}}

\address{Alexandra M. Schmidt\\
Estelina S. Capistrano \\
Departamento de M\'etodos Estat\'isticos \\
Instituto de Matem\'atica - UFRJ \\
Caixa Postal 68530 \\
Rio de Janeiro - R.J. Cep.: 21945-970 \\
Brazil \\
\printead{e1}\\
\phantom{E-mail:\ }\printead*{e3} \\
\printead{u1}}

\address{Marco A. Rodr\'iguez\\
D\'epartement des sciences de l'environnement \\
Universit\'e du Qu\'ebec \`a Trois-Rivi\`eres \\
3351, boulevard des Forges \\
Trois-Rivi\`eres, Qu\'ebec, G9A 5H7 Canada 
\\
\printead{e2}\\
\printead{u2}
}
\end{aug}

\renewcommand{\abstractname}{}

\begin{abstract}
Ecologists often interpret variation in the spatial distribution of populations in terms of responses to environmental features, but disentangling the effects of individual variables can be difficult if latent effects and spatial and temporal correlations are not accounted for properly. Here, we use hierarchical models based on a Poisson log-normal mixture to understand the spatial variation in relative abundance (counts per standardized unit of effort) of yellow perch, \emph{Perca flavescens}, the most abundant fish species in Lake Saint Pierre, Quebec, Canada. The mixture incorporates spatially varying environmental covariates that represent local habitat characteristics, and random temporal and spatial effects that capture the effects of unobserved ecological processes. The sampling design covers the margins but not the central region of the lake. We fit spatial generalized linear mixed models based on three different prior covariance structures for the  local latent effects: a single Gaussian process (GP) over the lake, a GP over a circle, and independent GP for each shore. The models allow for independence, isotropy, or nonstationary spatial effects. Nonstationarity is dealt with using two different approaches, geometric anisotropy, and the inclusion of covariates in the correlation structure of the latent spatial process. 
The proposed approaches for specification of spatial domain and choice of Gaussian process priors may prove useful in other applications that involve spatial correlation along an irregular contour or in discontinous spatial domains.
\end{abstract}


\end{frontmatter}

\section{Introduction}

\subsection{Ecological motivation\label{sec:motivation}}

Ecologists often seek to interpret variation in the spatial distribution of populations in terms of responses to environmental features. However, population distributions are influenced simultaneously by numerous environmental variables which vary in space and time and can not always be directly observed. Disentangling the individual effects of these variables on populations and, more generally, interpreting environmental effects in ecological analyses, can be difficult if the influence of latent (unobserved) variables and spatial and temporal correlations are not accounted for properly. Our aim in this study is to understand the spatial variation in relative abundance (counts per standardized unit of effort) of yellow perch, \emph{Perca flavescens}, the most abundant fish species in Lake Saint Pierre, Quebec, Canada. To this end, we use a hierarchical modelling approach to incorporate spatially varying environmental covariates that represent local habitat characteristics, and random spatial and temporal effects to capture the effects of unobserved ecological processes. Hierarchical modelling has proven to be a powerful tool for dealing with spatio-temporal variation and latent effects and attaining improved inference on specific environmental effects (reviewed in \citealp{wikle:2003}; \citealp{clark:gelfand:2006}; \citealp{wikle:2010}). To account for the spatial arrangement of the sampling locations and the pronounced heterogeneity of environmental influences across the lake, we explore various alternative specifications of the spatial domain and the spatial correlation structure.

\subsection{Geostatistical models for counts: a brief overview\label{sec:counts}}

Animal counts are often highly variable in space and time and show overdispersion relative to the Poisson distribution. Poisson-lognormal mixtures naturally incorporate overdispersion \citep{bulmer:1974}, and have been used to extend the conventional geostatistical framework to spatially structured counts: $Y({\bf s})$ is assumed to follow a conditional independent Poisson distribution with mean $\lambda({\bf s})$, where $\log \lambda({\bf s})={\bf X}({\bf s}){\bfbeta} +Z({\bf s})$, ${\bf X}({\bf s})$ is a $p$-dimensional row vector of covariates, ${\bfbeta}$ is the associated parameter column vector, and $Z({\bf s})$ is a local random effect \citep{diggle:moyeed:tawn:1998}. $Z(\cdot)$ is assumed to follow a zero-mean Gaussian process, with common variance $\sigma^2$ and valid correlation function $\rho({\bf s}-{\bf s}')$, where ${\bf s}$ and ${\bf s}'$ are arbitrary locations in the study region \citep{diggle:moyeed:tawn:1998}. The spatial process $Z$ is stationary if the correlation function $\rho(\cdot)$ is a function of the difference ${\bf s}-{\bf s}'$, and is both stationary and isotropic under the stronger assumption that $\rho(\cdot)$ is a function of the Euclidean distance between locations; the latter assumption implies that the distribution of $Z$ is invariant under translation and rotation. However, the distribution of $Y(\cdot)$ may not be stationary and isotropic even if $\rho(\cdot)$ is a function only of the Euclidean distance between locations. The vector defined as ${\bflambda}=(\lambda({\bf s}_1),\cdots,\lambda({\bf s}_n))'$ follows a multivariate log-normal distribution, and the marginal moments of the process $Y(\cdot)$ are obtained through the properties of the log-normal distribution and conditional expectations, that is
\begin{eqnarray*}
E(Y({\bf s}))&=&\exp(\mu({\bf s})+\sigma^2/2)=\alpha({\bf s}) \\
Var(Y({\bf s}))&=&\alpha({\bf s})+\alpha^2({\bf s})\left[\exp(\sigma^2)-1\right] \\
Cov(Y({\bf s}),Y({\bf s}'))&=&\alpha({\bf s})\alpha({\bf s}')\left[\exp(\sigma^2 \rho({\bf s}-{\bf s}'))-1 \right],
\end{eqnarray*}
where $\mu({\bf s})={\bf X}({\bf s}){\bfbeta}$. If one of the elements of ${\bf X}(\cdot)$ varies with location, then the resultant covariance function for $Y(\cdot)$ will also vary with location.

Although the inclusion of spatially varying covariates can yield an anisotropic process for $Y(\cdot)$, it may still be useful to consider models allowing {\em a priori} for anisotropy in spatial process $Z$ (\citealp{denison:mallick:1998}; \citealp{williams:1998}). However, the performance of these models would depend critically on whether the likelihood contains information to support inference on the parameters characterizing the anisotropy \citep{diggle:moyeed:tawn:1998}. Various geostatistical approaches to modelling nonstationary continuous data, including the use of covariates in the dependence structure are reviewed in \cite{guttorp:schmidt:2013}. Recent applications that use covariates in non-stationary dependence structures include \cite{ingebrigtsen:lindgren:steinsland:2014} and \cite{poppick:stein:2014}. Inclusion of covariates in the dependence structure can be a parsimonious way of capturing the effect of latent processes that modify the effective distance between points, such as physical mechanisms that facilitate or hinder the transport of materials or energy between points. 
In the absence of detailed subject knowledge of the mechanisms behind spatial connections, the covariates in the dependence structure can provide a simple proxy for those mechanisms, and their interpretation may yield insight into potential sources of anisotropy in the study system.

\subsection{Sampling design and data collection\label{sec:sampling}}

\begin{figure}[!bht]
\begin{center}
\subfigure{\includegraphics[width=0.5\textwidth]{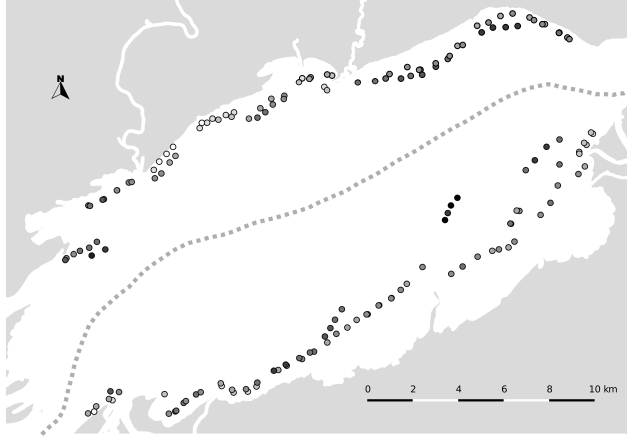}} 
\caption{Study locations along the north and south shores of Lake Saint Pierre (circles). Each location represents the centroid of a fishing trajectory approximately 650 m in length. Symbol shading is proportional to geodetic lake depth. The central navigation channel is represented by the dotted curve. Water flows from south-west to north-east. \label{fig:LSP}}
 \end{center}
\end{figure}

\paragraph{Study system\label{sec:sampling}}
Fish counts and environmental measurements were collected between 14 June and 22 August 2007 in Lake Saint Pierre (46$^\circ$12\textsf{'} N; 72$^\circ$50\textsf{'} W), a fluvial lake of the Saint Lawrence River (Quebec, Canada). Environmental conditions in the lake show strong spatial heterogeneity and temporal variability. The lake is large (surface area: annual mean  315 km$^2$; 469 km$^2$ during the spring floods) and shallow (mean water depth  3.17 m; range  1.23 m). The ice-free (April-November) surface area of LSP fluctuates between 387 and 501 km$^2$ depending on water level (\citealp{hudon:1997}). Lake Saint Pierre has distinct water masses running along its northern, central, and southern portions. These water masses originate from different sources and differ consistently in physical and chemical characteristics because lateral mixing is limited. A deep ($>$14 m) central navigation channel runs along the major axis of the lake; the channel has strong current and may act as a barrier to fish movement between the north and south shores (Fig. \ref{fig:LSP}).

\paragraph{Fish relative abundance\label{sec:sampling}}
Fish were collected from the shallow littoral zone ($< 2.5$ m depth) by electrofishing (Smith-Root CataRaft boat). Counts of yellow perch  were obtained for $n=160$ locations equally distributed between the north and south shores of the lake (Fig. \ref{fig:LSP}). Each location represents the centroid of a 20-min fishing trajectory measuring approximately 4 m in width and 650 m in length and running parallel to the shoreline. The fishing trajectories provided extensive coverage of the available littoral habitat along both shorelines. All samples were collected by a single team of trained operators. Fish sampling was carefully standardized to reduce variation in sampling efficiency among locations: current intensity was always maintained between 6 and 7 amp to control for variation in water conductivity; fish were only collected near the water surface to reduce the effect of water transparency on visibility; operators were equipped with polarized sunglasses and visors to reduce glare and improve visibility; mean depth at all locations ($<1.56$ m) allowed for coverage by the electric field of the entire water column along the trajectory; sampling was only conducted on days with at most moderate breeze ($\leq4$ on the Beaufort scale). This protocol yielded a single measurement of relative abundance (counts per 20 min of standardized sampling) for each location. To reduce the time and effort required to move between locations, samples were collected from a cluster of either four or eight adjacent locations on each of 38 sampling dates (four locations on 36 dates; eight locations on two dates). The sampling dates were unevenly spaced in time over a period of 70 days, and the north and south shores were visited in alternation on consecutive sampling dates. This sampling design yielded measurements that were clustered both in space and in time, in contrast with the simultaneous sampling of all locations at all occasions characteristic of many spatio-temporal sampling schemes.

\paragraph{Environmental covariates for the Gaussian process mean structure\label{sec:sampling}}
In Lake Saint Pierre, suitable habitat for yellow perch is concentrated in the shallow littoral zones that border the lake shores. Fish counts are expected to respond locally to habitat conditions at the site of capture, which are characterized in this study by a set of four environmental covariates measured  at each location: water depth, transparency, vegetation, and substrate composition. These four covariates together with an intercept were included in the mean structure of all models considered here.

\paragraph{Geodetic covariate for the Gaussian process covariance structure\label{sec:sampling}}

Two of the models we consider include, in addition to the environmental covariates in the mean structure, a covariate in the correlation function, geodetic lake depth, which allows for non-stationarity of the covariance structure of the spatial random effects. Including information on this covariate in the covariance structure of the spatial process provides a flexible yet relatively simple means of capturing anisotropy along the shorelines of the lake. Geodetic lake depth, measured as water depth minus the lake level relative to a fixed International Great Lakes low-water datum (IGLD55), was calculated for each location and sampling date. Locations at a given geodetic depth lie along a common isobath, or equal-depth contour along the lake bottom (Fig. \ref{fig:LSP}). Relative abundance at sites distant from each other but having similar geodetic depth may be linked by
%current flow, shoreline-related processes such as duration and extent of inundations, or behavioural 
dynamic processes such as fish movements along lake depth contours. Large-scale processes such as movement can generate spatial correlation in fish counts, yet their effects may not be adequately captured by the local environmental covariates, which are typically selected to reflect habitat preference at smaller scales. When information on movement dynamics is not available for inclusion in the mean structure, inclusion of covariates such as geodetic depth in the covariance structure may be useful for capturing correlations generated by the underlying latent process.

\paragraph{}

This paper is organized as follows. Section \ref{sec:model} describes the spatial generalized linear mixed models (SGLM) considered in this study. We explore three classes of Gaussian processes for the local latent effects: one defined over the whole lake, another defined over a circle, and lastly one that assumes independent Gaussian processes for each shore. For the models that assume either a single Gaussian process over the lake or independent Gaussian processes for each shore, we also allow for anisotropy of the local effects. Anisotropy is represented either geometrically or, alternatively, by use of a covariate in the correlation structure of spatial processes. Section \ref{sec:likelihood} describes the inference and model selection procedures. Section \ref{sec:results} presents and interprets the results obtained under the fitted models, including the potential influence of spatial confounding. Section \ref{sec:conclusion} concludes by discussing the advantages of the fitted models for understanding the spatial distribution of yellow perch in Lake Saint Pierre.

\section{Modelling environmental covariate, temporal, and spatial effects\label{sec:model}}

We assume that observed fish counts are realizations from a Poisson-lognormal mixture. Specifically, $Y({\bf s})$ is the number of fish observed at location ${\bf s}$ and 
\begin{eqnarray*} 
Y({\bf s})\mid \lambda({\bf s}) \sim Poi(\lambda({\bf s}))
\end{eqnarray*}
follows a conditionally independent Poisson distribution.
We assume
\begin{eqnarray}
\log(\lambda({\bf s}))={\bf X}({\bf s}){\bfbeta}+\delta({\bf s}), \label{eq:covar}
\end{eqnarray}
where ${\bf X}({\bf s})$ is a $K$-dimensional row vector containing a value of $1$ and the $K-1$ environmental covariates observed at location ${\bf s}$ and ${\bfbeta}=(\beta_0,\beta_1,\cdots,\beta_{K-1})'$ is the corresponding parameter vector of regression coefficients (i.e., $\beta_0$ is an intercept). The second component in (\ref{eq:covar}), $\delta(\cdot)$, is a mixing component that comprises temporal and spatial random effects and allows for overdispersion in the Poisson distribution. 
We decompose this mixing component as the sum of two independent terms,
\begin{eqnarray}
\delta({\bf s})=\gamma(t({\bf s}))+Z({\bf s}) \, , \label{eq:mix}
\end{eqnarray}
where $\gamma(t({\bf s}))$ captures temporal effects common to the $n_t$ locations sampled on a particular day, and $t(\cdot)$ is a deterministic function that assigns to each location ${\bf s}$ the ordinal rank corresponding to the Julian day on which the location was sampled, that is, $\{t(\cdot) \in 1,\cdots,T\}$, where $T = 38$ is the total number of sampling days. The spatial term $Z({\bf s})$ captures local effects that remain after accounting for environmental covariate and temporal effects. The simplifying assumption of additivity of temporal and spatial terms was required because substantially more replication would have been needed to allow for inclusion of time-space interactions. In all, 11 models are considered, all of which include the four environmental covariates as well as temporal effects. As well, all models but one (a non-spatial``baseline" model; see below) include spatial effects.

\subsection{Modelling the temporal effect $\gamma(t(\cdot))$}

Sampling days are unevenly spaced. For each sampling day $t$ we have observations at $n_t$ different locations. We initially assumed that the temporal effects ${\bfgamma}=(\gamma(1),\cdots,\gamma(T))'$ follow a multivariate zero-mean normal prior distribution and that the correlation between components of $\gamma$ decay exponentially in time, with $Cov(\gamma(t),\gamma(t'))=\tau^2 \exp\left(-\frac{1}{\phi_{\gamma}} \, |J(t)- J(t')|\right)$, where $J(t)$ is the Julian day associated with the $t$-th sampling day, $\tau^2$ represents the variance of $\gamma$, assumed constant across time, and $\phi_{\gamma}>0$ captures time-dependent decay in the correlation structure among elements of $\gamma$. Exploration of different prior specifications for $\phi_{\gamma}$ yielded no evidence of correlation in the components of ${\bfgamma}$ across time. We therefore retain a simpler independence structure in subsequent analyses; all results reported in Section \ref{sec:results} are based on an independent normal prior distribution with zero mean and variance $\tau^2$ for each $\gamma(\cdot)$.

\subsection{Modelling the local effects $Z(\cdot)$} %\label{sec:priorS}}

Let ${\bf Z}=(Z({\bf s}_{1}),\cdots,Z({\bf s}_{n}))'$ be the $n$-dimensional vector obtained by stacking the latent local effects of all observed locations $Z({\bf s}_i)$, $i=1,\,\cdots,n$. We assume that ${\bf Z}$ is a partial realization from a zero-mean Gaussian process with covariance matrix ${\bfSigma}$. A key aspect differentiating the models we explore here is the definition of the covariance matrix ${\bfSigma}$. To account for spatial effects arising from unmeasured (latent) ecological processes, such as land-water exchanges, barrier effects of the central navigation channel, or behavioural aggregation, we examine 10 spatial models representing different combinations of spatial domain and spatial correlation structure. For comparison we include as well a baseline model with no spatial effects (M0), that is $Z({\bf s})=0, \forall {\bf s}$. The spatial domain and correlation structure of the 11 models under consideration, labelled M0 through M10, are summarized in Table \ref{tab:models}.

\begin{table}[!htb]
\caption{Spatial domain and correlation structure of the 11 models considered in this study. \label{tab:models}}
\begin{center}
\begin{tabular}{cccccc}
\hline 
\multirow{1}{*}{Spatial} & \multicolumn{5}{c}{\multirow{1}{*} Correlation structure} \\
domain & & & & & \\
\cline{2-6}  \multirow{1}{*} & None \multirow{1}{*} & Independence & Isotropy & \multicolumn{2}{c}{\multirow{1}{*}Anisotropy} \\
\cline{5-6}  & & & & Geometric & Covariate \\
\multirow{1}{*} & & & & & in \\
\multirow{1}{*} & & & & & correlation \\
\hline  
\vspace{0.05in}
 & M0 &  &  &  \\
Whole lake & & M1 & M2 & M3 & M4 \\
\vspace{0.05in}
Circular & & & M5, M6 & & \\
By shore & & M7 & M8 & M9 & M10 \\
\hline
\end{tabular}
\end{center}
\end{table}

\subsubsection{Specification of spatial domain and choice of Gaussian process priors}

\paragraph{A single Gaussian process over the lake (models M1-M4)}

As a first approximation, the components of ${\bf Z}$ may be viewed as a partial realization from a random field defined over the whole lake. We assume that the local effects ${\bf Z}$ follow a zero-mean Gaussian process prior such that
${\bf Z} \sim N(0,{\bfSigma})$, where ${\bfSigma}$ is a $n$-dimensional covariance matrix, with each element given by $\Sigma({\bf s},{\bf s}')=\sigma^2 \, \rho({\bf s},{\bf s}';\phi)$, where $\sigma^2$ is the variance of the process and $ \rho({\bf s},{\bf s}';\phi)$ is a valid correlation function that depends on a parameter vector $\phi$. The hyperparameters to be estimated for this Gaussian process are ${\bfeta}^Z=(\sigma^2,\phi)$.

\paragraph{A single Gaussian process over a circle (models M5 and M6)}

The sampling locations form an approximately elliptical arrangement along the north and south shorelines of the lake (Fig. 1a of Section 1 of the supplemental article \citep{schmidt:rodriguez:capistrano}). Correlations between locations may therefore be induced by ecological processes associated with depth contours that follow the shoreline, such as water flow and fish movements. We examine this possibility by projecting the locations onto a unit circle and fitting models that assume a Gaussian process over a circle. Let $\omega$ be the angular distance between any two points $c_1$ and $c_2$ on the circle $\mathbb{S}$. The Euclidean distance between points $c_1$ and $c_2$ (i.e., the chord distance) is $\xi = 2\,r\sin(\omega/2)$. The function $\rho_\xi(\omega;\phi)=\exp\left\{-\frac{1}{\phi} \, \, 2\,r\sin(\omega/2)\right\}$ is a possible correlation function for a homogeneous random field on the circle $\mathbb{S}$ \citep[pp. 387-389]{yaglom:1987}. \cite{gneiting:2013} notes that the functions $\rho_\xi(\omega;\phi)$ and $\rho_\omega(\omega;\phi)=\exp(-\omega/\phi)$, for $\omega \in [0,\pi]$, are both valid, positive definite correlation functions on the circle, and have similar behaviour, particularly in the critical neighbourhood of $\omega = 0$. Here, we define an isotropic random field on a circle of radius $r=1$ and fit the circular models under both correlation functions, $\rho_\xi$ and $\rho_\omega$. The two correlation functions yielded very similar results and therefore only those based on $\rho_\xi$ are presented in Section \ref{sec:results}. The hyperparameters to be estimated for the circular Gaussian process are ${\bfeta}^{\mbox{\footnotesize{Circ}}}=(\sigma^2,\phi)$.

Two different circular projections of the sampling location coordinates are considered: 
\begin{enumerate}
\item Model M5: An ellipse is fitted to the UTM location coordinates by orthogonal least-squares (Fig. 1a  in Section 1 of the supplemental article \citep{schmidt:rodriguez:capistrano}). The original space is then shrunk to yield identical ellipse semi-axes of unit length, and the shrunken location coordinates are projected radially (scaling by vector norm) onto the resulting unit circle (Fig. 1b  in Section 1 of the supplemental article \citep{schmidt:rodriguez:capistrano}); 
\item Model M6: The centred location coordinates are projected radially (scaling by vector norm) onto a unit circle (Fig. 1c in Section 1 of the supplemental article \citep{schmidt:rodriguez:capistrano}).
\end{enumerate}
Note that the assumption of isotropy of the local effects holds only after the projection of the original geographical locations onto a circle.

\paragraph{Separate Gaussian processes for the north and south shores (models M7-M10)}
The potential of the navigation channel to act as a barrier to ecological exchanges between shores, as well as previous work which points to marked differences in the spatial structure of yellow perch growth between the two shores \citep{glemet:rodriguez:2007} suggest that local effects may have different covariance structures in the north and south shores. We therefore explore models that assume independent zero-mean Gaussian processes over each shore, each with its own covariance structure: ${\bfSigma}_N({\bf s},{\bf s}')=\sigma^2_N\, \rho({\bf s},{\bf s}';\phi_N)$ for the north shore and ${\bfSigma}_S(s,s')=\sigma^2_S \, \rho({\bf s},{\bf s}';\phi_Z)$ for the south shore.

\subsubsection{Specification of spatial correlation structures: isotropy and nonstationary spatial effects\label{sec:correlation}} 
For the isotropic cases we assume an exponential correlation function, $\rho(d,\phi)=\exp \left(-\frac{1}{\phi} ||{\bf s}-{\bf s}'|| \right)$, where $d=||{\bf s}-{\bf s}'||$ denotes Euclidean distance between locations ${\bf s}$ and ${\bf s}'$, and $\phi>0$ is the decay parameter of the correlation function. The hyperparameters to be estimated in the isotropic covariance structure are $\eta^I=(\sigma^2,\phi)$.

In the present study water flow, lake morphometry, land-water exchanges, and the presence of the central navigation channel may all induce directional effects on the correlations between fish counts. We therefore explore correlation functions for the local spatial effects ${\bf Z}$, that relax the assumption of isotropy. Specifically, we discuss two different approaches, geometric anisotropy, and inclusion of covariates in the correlation function.

\paragraph{Geometric anisotropy (models M3 and M9)} In this approach, a stationary covariance structure is transformed by differential stretching and rotation of the coordinate axes to capture directional effects  \citep{diggle:ribeiro:2007}. In the models considered here, the spatial effects ${\bf Z}$ are correlated as a function of a linear transformation of the original coordinate system. We assume ${\bfSigma}({\bf s},{\bf s}')=\sigma^2 \rho(||f({\bf s})-f({\bf s}')||,\phi)$, where $f({\bf s})={\bf s}\,{\bf A}$, and 
$$
{\bf A}=\left[
\begin{array}{cc}
\cos \psi_A & -\sin \psi_A \\
\sin \psi_A & \cos \psi_A \\
\end{array}\right]
\left[
\begin{array}{cc}
1 & 0 \\
0 & \psi_R^{-1}
\end{array}\right],
$$ 
where $\psi_A \in (0,2\pi)$ is the anisotropy angle and $\psi_R>1$ is the anisotropy ratio \citep{diggle:ribeiro:2007}. The hyperparameters to be estimated in the geometric anisotropy model are ${\bfeta}^G=(\sigma^2,\phi,\psi_A,\psi_R)$.

\paragraph{Inclusion of covariates in the spatial correlation function (models M4 and M10)} A potential limitation of geometric anisotropy models is that they rely on a highly symmetrical representation that can describe global directional features over the study region, but not local patterned variation. An alternative approach, which retains model simplicity while affording additional flexibility, is to allow for inclusion of covariates in the correlation structure of Gaussian processes. Following \cite{schmidt:rodriguez:2011a}, we include geodetic depth as a covariate in the correlation structure of ${\bf Z}$ to allow for non-stationarity of the resultant covariance structure of the local random effects. Specifically, we assume ${\bf s}_c=(s_1,s_2,q(s_1,s_2))'=({\bf s},q({\bf s}))'$, where $s_1$ and $s_2$ are easting and northing coordinates and $q(s_1,s_2)$ is the observed geodetic depth at location ${\bf s}=(s_1,s_2)$. The covariate $q(.,.)$ is a function of $s_1$ and $s_2$, and so one can think of this spatial process as defined in a two-dimensional manifold \citep{schmidt:guttorp:ohagan:2011}. We model the elements of ${\bfSigma}$ as 
\begin{eqnarray*}
{\bfSigma}({\bf s}_c,{\bf s}_c')=\sigma^2 \exp \left\{-\frac{1}{\phi_1} \mid \mid {\bf s}-{\bf s}' \mid \mid -\frac{1}{\phi_2} |q({\bf s})-q({\bf s}')|\right\},
\label{eq:mahalanobis}
\end{eqnarray*}
with $\phi_1,\phi_2>0$. This covariance function is non-stationary in the two-dimensional manifold. For this model, the hyperparameters to be estimated are ${\bfeta}^C=(\sigma^2,\phi_1,\phi_2)$.

\section{Inference procedure and model comparison\label{sec:likelihood}}

\paragraph{Likelihood function}
Let ${\bf y}=(y({\bf s}_1),\cdots,y({\bf s}_n))'$ be the observed count at each of the sampling locations (Fig. \ref{fig:LSP}) over the sampling period. The vector ${\bftheta}$ comprises all the parameters and hyperparameters involved in the model. The hyperparameters to be estimated are those related to the prior specification for ${\gamma}(.)$ and ${\bf Z}$ as discussed above. Conditional on $\lambda({\bf s}_i)$ each observation is an independent realization from a Poisson distribution; therefore, the likelihood function follows the relationship 
\begin{eqnarray*}
f({\bf y}\mid {\bftheta}) \propto \prod_{i=1}^{n }
\exp\left\{-\lambda({\bf s}_i) \right\} \,
\left[\lambda({\bf s}_i)\right]^{y({\bf s}_i)}.
\end{eqnarray*}
We follow the Bayesian paradigm to obtain estimates of the unknowns in the model. 

\paragraph{Prior specification of the components of ${\bftheta}$} We assume that all components of ${\bftheta}$ are independent {\em a priori}. For the components of ${\bfbeta}$ we assume a  normal prior distribution with zero mean and some large variance, to represent our lack of information on how each component of ${\bf X}(.)$ influences the mean of the Poisson distribution. For the variances $\tau^2$ and $\sigma^2$ we assume inverse gamma distributions with infinite variance and mean based on the residual standard error estimate of a log-linear fit. For the decay parameters of the correlation functions ($\phi$) we assign gamma prior distributions having unit mean and variance. The components of ${\bf Z}$ were not expected {\em a priori} to show very strong spatial correlation, and so the prior specification for $\phi$ sets the range of an isotropic spatial process at half or less of the maximum observed distance, with 99\% probability.

For the geometric anisotropy model we assume, {\em a priori}, $\psi_A \sim U(0,\pi)$ for identifiability (to ensure that orientations are unique within the interval considered). Prior specifications for the anisotropy ratio hyperparameter must satisfy the constraint $\psi_R\geq 1$. We assign a Pareto prior distribution to $\psi_R$ with scale parameter $1$ and shape parameter $2$. This prior has mode $1$ (corresponding to the isotropic case),  mean $2$, and infinite variance.

\paragraph{Posterior distribution} Following the Bayesian paradigm, the posterior distribution is proportional to likelihood times prior. For all models considered above, the posterior distribution does not admit a simple closed form. Therefore, we used Markov chain Monte Carlo algorithms, specifically Gibbs sampling with some Metropolis-Hastings (M-H) steps \citep{metropolis:rosenbluth:rosenbluth:teller:teller:1953,hastings:1970}, to sample from the posterior.

We reparametrized the model described in equations (\ref{eq:covar}) and (\ref{eq:mix}) to build a more efficient MCMC sampling scheme. We let $\log(\lambda({\bf s}_{i}))={\bf X}_i^*({\bf s}_{i}){\bfbeta}^*+W({\bf s}_{i})$, and $W({\bf s}_{i})=\beta_0+\gamma(t({\bf s}_i))+Z({\bf s}_{i})$, where ${\bf X}_i^*(.)$ does not have a column of ones and ${\bfbeta}^*=(\beta_1,\cdots,\beta_{K-1})$. The covariate coefficients do not have full conditional posteriors of closed form, so they were sampled using M-H steps whose proposal distribution was based on the algorithm proposed by \citet{gamerman:1997}. The $W({\bf s}_{i})$ were sampled using a random walk M-H step with the proposal variance tuned to yield acceptance rates approaching 0.44 \citep{roberts:rosenthal:2009}. Let ${\bf W}$ and ${\bf Z}$ be the vectors obtained by separately stacking the $W({\bf s}_{i})$ and $Z({\bf s}_{i})$. ${\bf Z}$ is assumed to follow a zero-mean multivariate normal distribution, with covariance matrix ${\bfSigma}$. We can write $ {\bf W}={\bf 1}_n \beta_0 + {\bf B}{\bfgamma}+{\bf Z}$, which follows an $n$-dimensional multivariate normal distribution, with mean $ {\bf 1}_n \beta_0 + {\bf B}{\bfgamma}$, and covariance matrix ${\bfSigma}$. Here, ${\bfgamma}=(\gamma(1),\cdots,\gamma(T))'$ is the $T$-dimensional vector comprising the temporal random effects, and ${\bf B}$ is a $n \times T$ matrix, where each row is given by the $T$-dimensional row-vector ${\bf e}_t$ having $t^{th}$ column equal to 1 and all other elements equal to zero. Row vector ${\bf e}_i$ enters in ${\bf B}$ $n_t$ times. Under this reparametrization, it is easy to sample from the known full conditional posteriors of $\beta_0$ and ${\bfgamma}$ (normal distributions), and of $\sigma^2$ and $\tau^2$ (inverse-gamma distributions). The parameters in the spatial correlation function result in unknown full conditional posteriors from which we sampled using M-H steps. For all models, we ran the MCMC algorithm (two chains with overdispersed starting values) for 70,000 iterations, used 10,000 iterations as burnin, and stored every 60$^{th}$ iteration. The MCMC algorithm was implemented in the Ox programming language, v. 6.20 \citep{doornik:2007}. Convergence was checked using the diagnostics in {\tt R} package {\tt coda} \citep{plummer:2006}.

\paragraph{Model comparison} Four different criteria were used to compare models: (i) deviance information criterion (DIC) \citep{spiegelhalter:best:carlin:linde:2002}, (ii) ranked probability score (RPS), (iii) logarithmic score (LS), and (iv) Dawid-Sebastiani score (DSS) (Section 2 of the supplemental article \citep{schmidt:rodriguez:capistrano}). The last three criteria are proper scoring rules, proposed by \cite{gneiting:raftery:2007} and discussed for discrete observations in \cite{czado:gneiting:held:2009}. As emphasized by \cite{czado:gneiting:held:2009}, propriety is an essential property of a scoring rule that encourages honest and coherent predictions and ensures that both calibration (consistency between predictions and observations) and sharpness (concentration of predictive distributions) are being addressed. Following \cite{gsch:czad:2008}, we use the same data for estimation and computation of the scores, as our focus is on understanding the distribution of the counts over the lake rather than prediction. For all criteria, the best model among those fitted is that having the smallest value for the criterion.

\section{Results\label{sec:results}}
\paragraph{Model comparisons based on DIC, RPS, LogS, and DSS}
The baseline model excluding spatial effects (M0) had very poor performance relative to all other models (Table \ref{tab:comparison}). Comparisons among the models that assume a single Gaussian process over the lake (M1 to M4), and among those that allow for different local spatial processes on the north and south shores (M7 to M10), indicate that anisotropic models provided better fits than the independent or isotropic models (Table \ref{tab:comparison}). Overall, the model that performed best under all four criteria (M9) incorporated both anisotropy and a separate Gaussian process for each shore. Posterior predictive checks for M9 (Fig. 2 in Section 3 of the supplemental article \citep{schmidt:rodriguez:capistrano}) show good agreement between observed and fitted values. The importance of considering separate spatial structures for the north and south shores is also supported by the fact that M5, which emphasizes the distance between shores relatively more than M6, performs better than the latter under all criteria. These results suggest that after accounting for the effects of time and of the measured environmental covariates, the spatial distribution of yellow perch in Lake Saint Pierre is better modelled with local random effects that originate from different processes in the north and south shores. 

\begin{table}[!htb]
\caption{Model comparisons based on DIC, RPS, LogS, and DSS. For each criterion, the value associated with the best-performing model is given in bold characters.\label{tab:comparison}}
\begin{center}
\begin{tabular}{lccccccc}\hline
Model & Correlation structure & $\overline{D}$ & $p_D$ & DIC &  RPS & LogS & DSS \\ \hline
M0 & None  & 1591.7 & 36.6 & 1628.2 & 6.42 & 4.98 & 25.2 \\ \hline
\multicolumn{7}{c}{Whole lake} \\ \hline
M1 & Independence & 905.0 & 133.6 & 1038.6 &  2.56 & 2.83 & 3.59 \\
M2 & Isotropy & 906.9 & 127.2 & 1034.1 &  2.60 & 2.83 & 3.58 \\
M3 & Anisotropy - Geom. & 906.2 & 127.2 & 1033.4 &  2.61 & 2.83 & 3.58 \\
M4 & Anisotropy - Cov. in cor. & 902.9 & 128.4 & 1031.3 &  2.57 & 2.82 & 3.57 \\ \hline
\multicolumn{7}{c}{Circular} \\ \hline
M5 & Isotropy & 910.1 & 131.0 & 1041.0 &  2.60  & 2.84 & 3.60 \\
M6 & Isotropy & 918.5 & 133.1 & 1051.6 &  2.68 & 2.87 & 3.63 \\ \hline
\multicolumn{7}{c}{By shore} \\ \hline
M7 & Independence & 909.0 & 133.4 & 1042.4 &  2.58 & 2.84 & 3.59 \\ 
M8 & Isotropy & 899.3 & 126.5 & 1025.8 &  2.56 & 2.81 & 3.54 \\
M9 & Anisotropy - Geom. & 897.3 & 125.1 & {\bf 1022.4} &  {\bf 2.54} & {\bf 2.80}  & {\bf 3.54} \\
M10 & Anisotropy - Cov. in cor. & 900.8 & 129.1 & 1029.9 &  2.55 & 2.81 & 3.56\\\hline 
\end{tabular}
\end{center}
\end{table}

In what follows we focus on results obtained under the four models that assume different local structures of the covariance matrix ${\bfSigma}$ for the north and south shores (M7, M8, M9, and M10). Note that these four models can be viewed as nested: M8 is a particular case of M9 if $\psi_A=0$ and $\psi_R=1$, M10 is in turn a particular case of M8 if $\phi_2\rightarrow \infty$, and M7 is a particular case of M8, M9, and M10, obtained by setting the appropriate decay parameters $\phi \approx 0$. For this reason, we examine the posterior distributions of the parameters to assess the information gain relative to the prior under each of these models, as suggested by \cite{schmidt:rodriguez:2011b}.

\begin{figure}[!h]
\begin{center}
\includegraphics[scale=0.20]{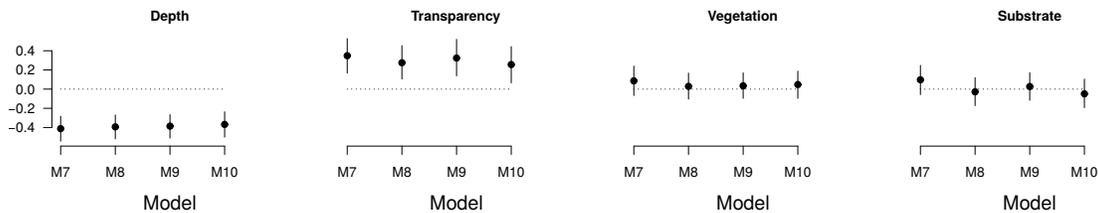}
\caption{Posterior summary of regression coefficients $(\beta_1,\beta_2,\beta_3,\beta_4)'$, for the four environmental covariates (depth, transparency, vegetation, and substrate) included in the mean structure of the model. Results are shown for the four models that assume different local structures of the covariance matrix $\Sigma$ for the north and south shores: M7, M8, M9, and M10. Solid circles: posterior mean of $\beta$; vertical lines: 95\% credible interval. 
\label{fig:beta}}
\end{center}
\end{figure}

\paragraph{Environmental effects}
The prior structure assumed for the latent local effects does not seem to influence much the estimates of the effects of the environmental covariates that determine the mean structure of the Gaussian process (Fig. \ref{fig:beta}). The responses to envrionmental covariates in models M7 through M10 suggest that the relative abundance of yellow perch in Lake Saint Pierre is responding primarily to transparency (positively) and depth (negatively), and not to vegetation or substrate. The estimated effects of transparency and depth are very substantial and thus have major implicatons for the spatial distribution of yellow perch in the lake. Over the observed ranges for these two covariates, these effects imply approximately four-fold (transparency) and eleven-fold (depth) changes in the Poisson mean for relative abundance.

\begin{figure}[!h]
\begin{center}
\includegraphics[angle=0,scale=0.3]{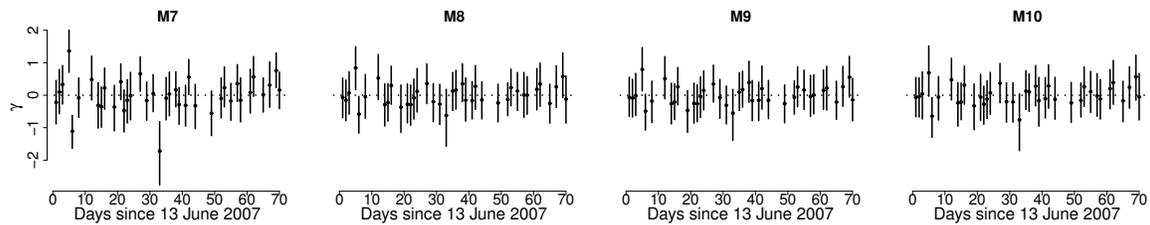}
\caption{Posterior summary of the temporal effects $\gamma(t({\bf s}_i))$,
for each of the $I=38$ sampling dates, under models M7, M8, M9, and M10.
Solid circles: posterior mean of $\gamma(.)$; vertical lines: 95\% credible interval. \label{fig:gammat}}
\end{center}
\end{figure}

\paragraph{Temporal effects}
Posterior distributions of temporal effects $\gamma(t(\cdot))$ under the independence model M7 seem to reflect structure in the data that is not apparent for the models that incorporate spatial correlation, M8, M9, and M10, none of which shows evidence of trends or other substantial temporal effects (Fig. \ref{fig:gammat}). M8, M9, and M10 show a generalized reduction in the variance of the temporal effects relative to M7, as seen in the reduced spread of those temporal effects most distant from zero, e.g., for days 5, 6, 27, 33, and 42. This reduction presumably reflects the use of information contained in spatial correlations to explain some of the variation attributed to the temporal effects in M7.

\begin{figure}[!h]
\begin{center}
\includegraphics[angle=90,scale=0.3]{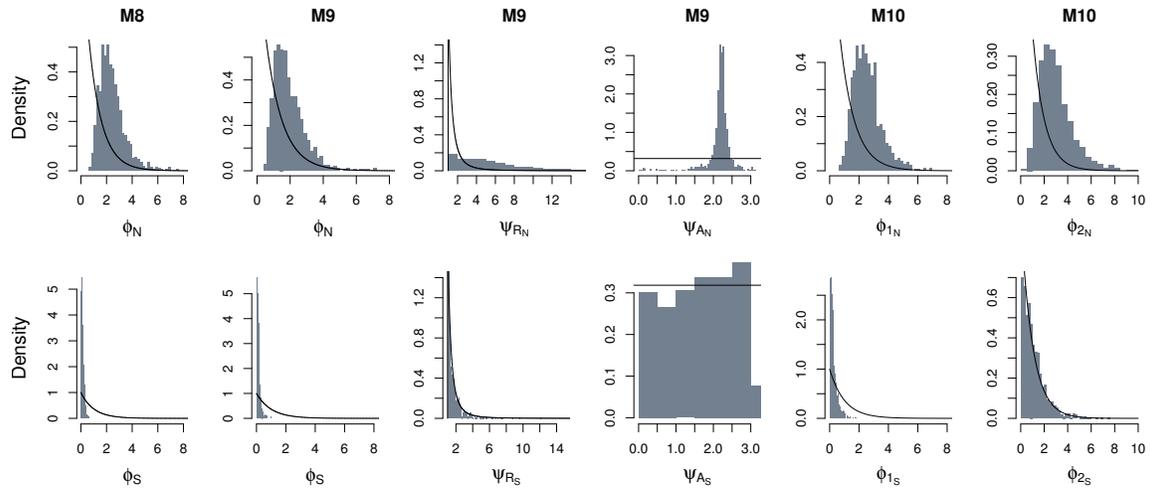}
\caption{Posterior distribution (histogram) and prior density (curve) of each of the hyperameters in the respective correlation functions of  models M8, M9, and M10 (columns), for the north and south shores (rows).
\label{fig:hyperAll}}
\end{center}
\end{figure}

\paragraph{Anisotropy}
Estimates of the decay parameters of the exponential correlation functions ($\phi_N$ and $\phi_S$) under M8, M9, and M10 suggest independence of the local effects on the south shore (second row of Fig. \ref{fig:hyperAll}). However, the posterior distributions for the anisotropy ratio ($\psi_R$) and anisotropy angle ($\psi_A$) under M9 provide strong evidence of anisotropy associated with the spatial process on the north shore, indicating a slower decay of correlation along the SW - NE direction (third and fourth columns of Fig. \ref{fig:hyperAll}). Strong evidence for anisotropy also emerges when geodetic depth is considered in the correlation structure of ${\bf Z}$ under M10 (fifth and sixth columns of Fig. \ref{fig:hyperAll}). The presence of spatial correlation in the north shore and apparent spatial independence in the south shore is consistent with the previous finding that individual growth of yellow perch shows marked spatial heterogeneity on the north shore, but spatial homogeneity on the south shore \citep{glemet:rodriguez:2007}.

\begin{figure}[!h]
\begin{center}
\subfigure[]{\includegraphics[scale=0.25]{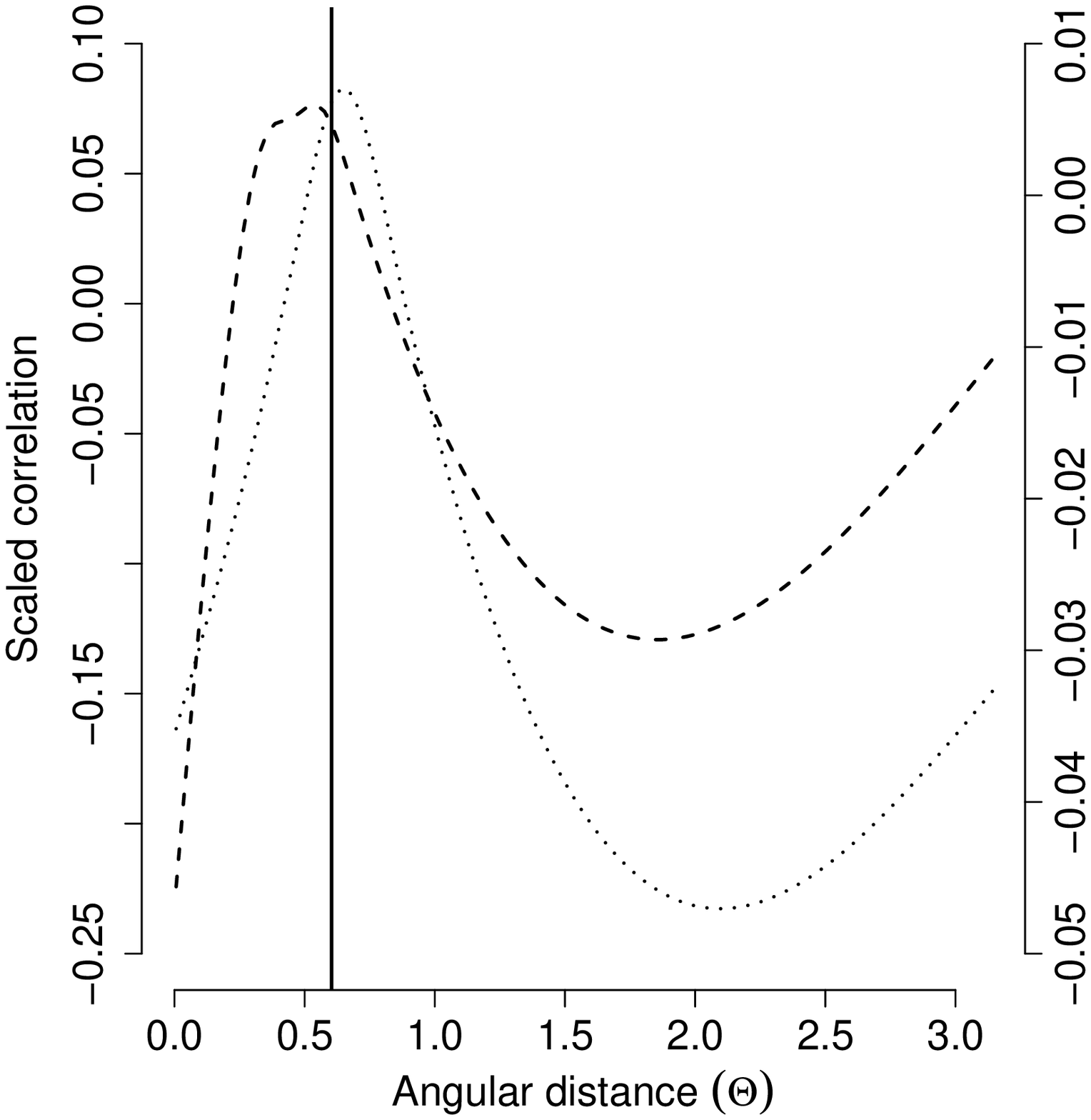}} \hspace{0.5cm}
\subfigure[]{\includegraphics[scale=0.25]{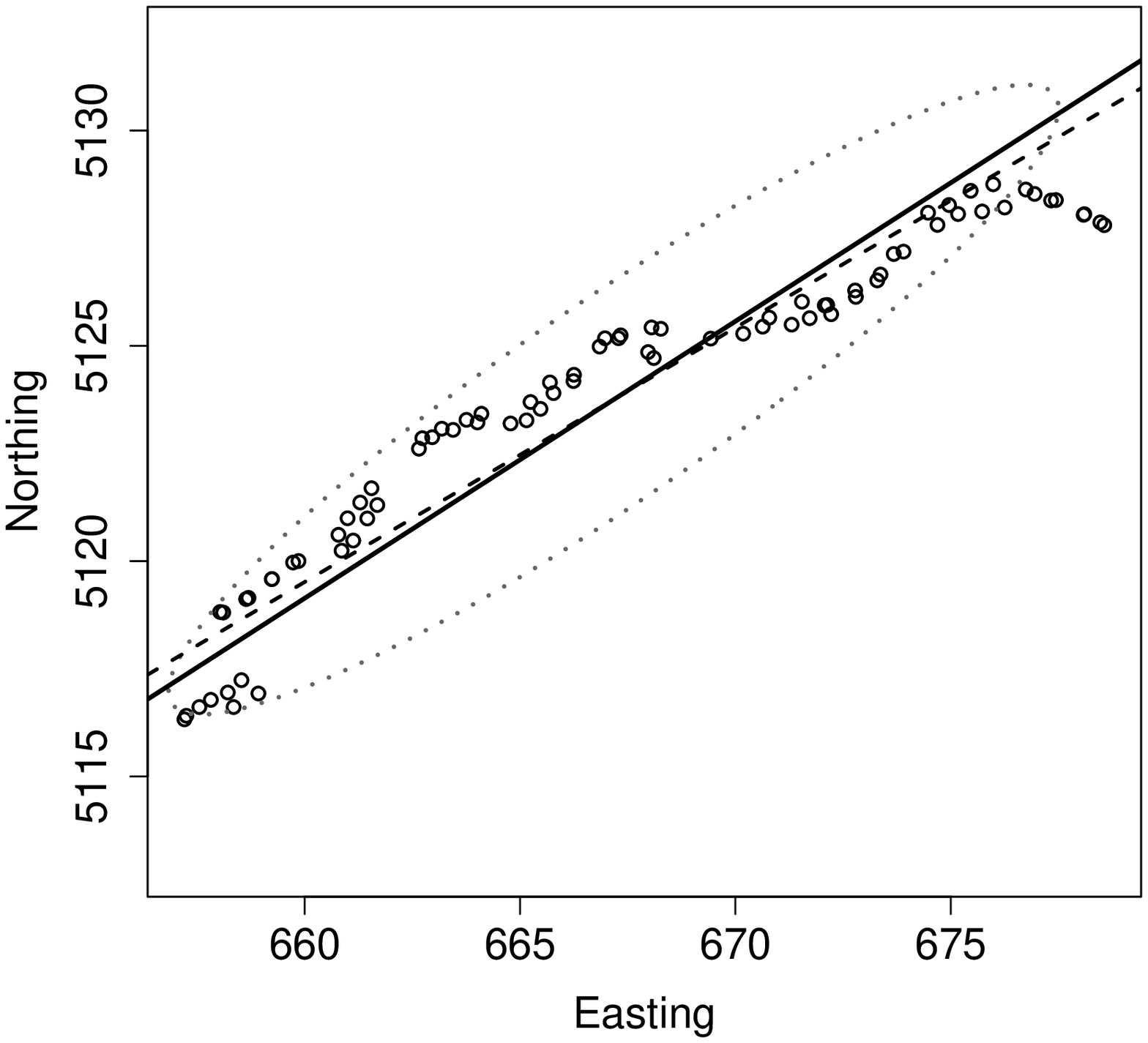}}
\caption{(a) Spatial correlation as a function of angular distance from the equator $(\Theta)$. Dashed curve: GAM smooth for the covariate-in-correlation model M10 (adjusted for geographical distance; y axis on the right). Dotted curve: GAM smooth for the geometric anisotropy model M9 (y axis on the left). Vertical line: posterior mean of the anisotropy angle $\psi_A$. (b) Anisotropic effects on the north shore. Sampling locations are shown together with 1) direction of lake major axis (solid line); 2) ellipse representing the anisotropy angle $\psi_A$ and ratio $\psi_R$ from model M9; 3) direction of least change in spatial correlation from model M10 (dashed line).}
\label{fig:gam}
\end{center}
\end{figure}

Geometric anisotropy can be readily represented graphically, e.g., by means of elliptical contours on a map. In contrast, the anisotropy estimated from covariate-in-correlation models can be difficult to visualize because it reflects underlying variation in the covariates, which may show little spatial pattern. Therefore, we used generalized additive model (GAM; \citealp{wood:2006}) analyses of spatial correlation as a heuristic device to understand the orientation of anisotropic effects in the covariate-in-correlation model M10. We calculated the pairwise geographical distances between all sampling locations and the smallest positive angle $\Theta$ subtended by the line connecting each pair of locations and the equator. A GAM was used to represent the spatial correlation fitted under the covariate-in-correlation model M10 as a function of the angle $\Theta$ and geographical distance. The results from this analysis were used to generate partial smooths showing change in spatial correlation as a function of angle $\Theta$ after adjustment for geographical distance. For comparison, this analysis was also performed for the spatial correlation fitted under the geometric anisotropy model M9. The GAM analyses for the covariate-in-correlation model M10 and the geometric anisotropy model M9 show close agreement and successfully retrieve the direction of least change in correlation specified by the fitted anisotropy angle $\psi_A$ (Fig. \ref{fig:gam}). Both anisotropic models indicate that spatial correlation decays most rapidly along a direction approximately perpendicular to the north shoreline and to the lake's major axis.

\paragraph{Latent spatial effects}
To better understand the behaviour of spatial effects on the north and south shores, we examined samples from the posterior distribution of the correlation among the components of ${\bf Z}_N$ and ${\bf Z}_S$. Models M8, M9, and M10 yield substantially different estimates for decay with distance (Fig. 3 top left in Section 3 of the supplemental article \citep{schmidt:rodriguez:capistrano}) and variability (Fig. 3 top right in Section 3 of the supplemental article \citep{schmidt:rodriguez:capistrano}) of spatial correlation among the local components on the north shore. Ranges of the posterior 95\% credible intervals of correlations under the geometric anisotropic model (M9) are much wider than those obtained under models M8 and M10; this may be related to the relatively uninformative prior assigned to $\psi_R$. In comparison to M8 and M9, M10 yields narrower ranges of the posterior 95\% credible intervals of correlations on the north shore (top right panel of Fig. 3 in Section 3 of the supplemental article \citep{schmidt:rodriguez:capistrano}). In contrast with the results for the north shore, models M8, M9, and M10 yield similar estimates of spatial correlation for the south shore, showing rapid decay of correlation with distance and providing another indication that the local latent effects are not spatially correlated on the south shore.

\begin{figure}[!h]
\begin{center}
\includegraphics[angle=0,scale=0.3]{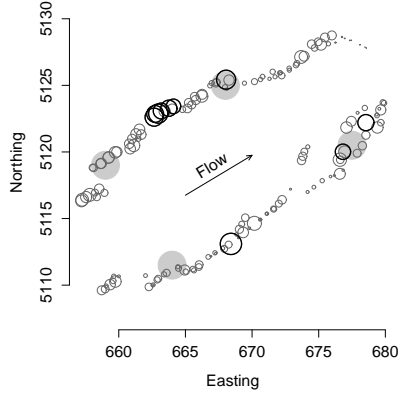} \\
\caption{Posterior mean of the local latent effects ${\bf Z}$ for the north and south shores under model M9 (open circles). The diameters of the open circles are proportional to $e^{Z({\bf s}_i)}$, $i=1,\cdots,n$. Locations corresponding to the 10 greatest posterior mean values of ${\bf Z}$ are shown as darker (black) circles. The approximate locations of the four known spawning areas in the lake are shown as shaded circles. \label{fig:SpatialEffects}}

\end{center}
\end{figure}

For the two shores, substantial spatial variation in relative abundance of yellow perch still remains after accounting for temporal and observed environmental effects (Fig. \ref{fig:SpatialEffects}). This local variation in relative abundance may be linked to unmeasured environmental effects, such as changes in optical, thermal, or chemical properties arising from the influence of tributaries that enter the lake. Local variation may also result from the concentration of mature adult fish at favorable spawning grounds during the reproductive season and subsequent downstream movement of groups of young fish from the spawning grounds. Larval fish have limited swimming ability and may be transported downstream by advection during the first weeks after spawning, until they have grown sufficiently to hold position against the flow. Interestingly, the strongest local latent effects appear to be concentrated in four areas located downstream of the four known spawning grounds identified in \cite{bertolo:2012}.

\paragraph{Spatial confounding}

A key objective in many applications involving spatial regression is to estimate the fixed effects while accounting for spatial correlation (\citealp{Reich:Hodges:Zadnik:2006}; \citealp{Hughes:Haran:2013}). Spatial confounding, which arises when the covariates and the spatial effects are not independent, can lead to estimates of the posterior mean and variance of fixed effects that markedly differ from those of the nonspatial regression model (\citealp{Reich:Hodges:Zadnik:2006}; \citealp{Paciorek:2010}; \citealp{Hodges:Reich:2010}; \citealp{Hughes:Haran:2013}). The most common approach for dealing with spatial confounding is restricted spatial regression (RSR), in which the spatial random effects are constrained to the orthogonal complement (residual space) of the fixed effects (\citealp{Hodges:Reich:2010}; \citealp{Hanks:Schliep:Hooten:Hoeting:2014}). The conditional likelihood functions of the data under SGLM and RSR are identical and therefore one can not choose between the two approaches based on in-sample data alone, which  can be problematic when the interpretation of fixed effects is of interest and the two approaches yield different estimates of these effects (\citealp{Hanks:Schliep:Hooten:Hoeting:2014}). 

In the geostatistical (continuous spatial support) setting, \citep{Hanks:Schliep:Hooten:Hoeting:2014} show that RSR provides computational benefits relative to the spatially confounded SGLM, but that Bayesian credible intervals under RSR can be inappropriately narrow under model misspecification. To mitigate this potential problem, they propose a posterior predictive approach (RSR-PPD) which adjusts the variance of the regressor coefficients to reflect possible collinearity between fixed and random effects.

To investigate possible spatial confounding between the covariates and the spatial effects $Z(\cdot)$, we adapted the restricted spatial regression (RSR) model proposed by \cite{Hanks:Schliep:Hooten:Hoeting:2014} to deal with a Poisson response variable, as detailed in Section 4 of the supplemental article \citep{schmidt:rodriguez:capistrano}. As all model comparison criteria pointed to the models with different spatial processes per shore as the best among those fitted under the SGLM approach (M8, M9, and M10; Table 2), we implemented the RSR and RSR-PPD approaches under the correlation structures corresponding to these models. We then compared the posterior summaries of the coefficients of the four environmental covariates under the SGLM (${\bfbeta}$), RSR (${\bfalpha}$), and RSR-PPD ($\tilde{\bfbeta}$) approaches. Similarly to \cite{Hanks:Schliep:Hooten:Hoeting:2014}, the posterior 95\% credible intervals of the covariate coefficients under RSR were narrower than those obtained under SGLM and RSR-PPD (Fig. 4 in Section 4 of the supplemental article \citep{schmidt:rodriguez:capistrano}). The intervals for vegetation and substrate overlapped zero under SGLM and RSR-PPD but were strictly positive under RSR (Fig. 4 in Section 4 of the supplemental article \citep{schmidt:rodriguez:capistrano}). The spatial covariance structure assumed for the spatial effect (M8, M9, or M10) did not seem to influence the posterior distribution of the coefficients. Interestingly, although the posterior means for the regressor coefficients under RSR and RSR-PPD were shifted relative to those under SGLM, substantive interpretation of covariate effects based on the 95\% credible intervals (i.e., whether intervals overlap zero) was similar under SGLM and RSR-PPM (only depth and transparency are important), and this interpretation contrasted with that under SRS (all covariates are important).

Following \citep{Hanks:Schliep:Hooten:Hoeting:2014} we also ran a simulation study to examine the influence of model misspecification on the coverage of covariate credible intervals. 
We generated multiple data sets from each of two models, a SGLM and a SRS, and then fit each model to all the data sets. For data generated from the SGLM, coverage of the true value of covariate coefficients was adequate for the SGLM and the RSR-PPD, but the RSR was unable to recover the true value, with the exception of transparency (Fig. 5 in Section 4 of the supplemental article \citep{schmidt:rodriguez:capistrano}). Conversely, for data generated from the RSR, coverage of the true value of covariate coefficients was similar for SGLM, RSR, and RSR-PPD (Fig. 5 in Section 4 of the supplemental article \citep{schmidt:rodriguez:capistrano}). In agreement with \citep{Hanks:Schliep:Hooten:Hoeting:2014}, we found that under model misspecification the covariate coefficients resulting from the RSR-PPD and SGLM approaches were conservative, whereas those from RSR could be inappropriately narrow.

Our results on spatial confounding are consistent with the caveat issued by \citep{Hanks:Schliep:Hooten:Hoeting:2014}, that when the generating mechanism for spatially autocorrelated observations is a spatially missing covariate, choosing the RSR over the SGLM assumes that this missing covariate is orthogonal to the measured covariates. Because smooth covariates are likely to be collinear, this generally is a strong assumption.

\section{Discussion\label{sec:conclusion}}

A hierarchical modelling approach was used to examine the variation in relative abundance of a fish species in a lake. We focused on a Poisson-lognormal mixture to model counts observed at locations along the shores of the lake over a 70-day period. We examined different candidate structures for the lognormal mixing structure, which include a temporal and a spatial component. The temporal component accounts for potential effects shared by locations sampled on the same day, whereas the spatial component accounts for effects arising from latent ecological processes. Environmental effects are incorporated by means of spatially varying covariates that reflect local habitat characteristics. In all, we examined 11 models which incorporated the same covariates and temporal effects but considered different combinations of spatial domain and spatial correlation structure (Table \ref{tab:models}). 

The local effects in models M5 and M6 are assumed to follow a Gaussian process over a circle once the sampling locations are rescaled by projection onto the unit circle. Gaussian processes defined on the circle have been used previously in biological applications, e.g., to describe the spread of an airborne plant disease from a point source \citep{soubeyrand:enjalbert:sache:2008}. This approach can be placed into the \cite{samp:gut:1992} framework, in which isotropy holds only after an unknown nonlinear transformation of the geographical locations. However, instead of seeking a nonlinear transformation to attain isotropy in the deformed space derived from the original configuration, we selected the circular transformations {\em a priori} based on ecological considerations. Specifically, the first projection (M5) reduces intra-shore distances relative to inter-shore distances and thereby emphasizes the potential for the central navigation channel to reduce inter-shore correlations, for example, by acting as a barrier to fish movement. In comparison, the second projection (M6) more closely approximates a scenario in which spatial correlations are determined by distance along the shoreline, with little influence of the navigation channel. This projection therefore emphasizes longitudinal effects related to water flow, such as contaminant or nutrient gradients along the plumes created by tributary streams and rivers entering the lake. The better performance of M5 relative to M6 points to the importance of inter-shore differences and hints at the operation of different ecological processes in the north and south shores. However, the poor performance of M5 and M6 relative to models that treat shores separately (Table \ref{tab:comparison}) suggests that the circular projection framework was not sufficiently flexible to capture differences between shores.

For the nested models M7-M10, which treat shores separately, comparisons of the prior-posterior information gain for the key parameters that differentiate the models (Fig. \ref{fig:hyperAll}) were a useful complement to the model comparison criteria when assessing the relative merits of the models. The inclusion of information on geodetic lake depth as a covariate in the covariance structure of the spatial process (M10) provided a flexible means of capturing anisotropy along the shorelines of the lake. However, the geometric anisotropy (M9) and covariate-in-correlation (M10) models yielded similar substantive results, presumably because the shape of the lake is compact and the arrangement of depth contours along both shores is regular. Spatial correlations induced by shoreline-related processes therefore have a simple structure that was captured adequately by either model.

Estimates of environmental effects are very similar for models M7-M10, showing little sensitivity to assumptions about the spatial term of the mixing component in those models. Among the environmental covariates considered, only water transparency and depth seem to influence the spatial distribution of yellow perch. The effects of transparency and depth on relative abundance of yellow perch are strong, which points to these water depth and transparency as potential determinants of the spatial distribution of this species in the lake. This finding agrees with earlier biological studies of local habitat preferences in yellow perch, which are reported to be most abundant in shallow, open waters of clear lakes with moderate vegetation and relatively fine substrate (silt to gravel) \citep{scott:crossman:1973}. However, we find no effect of vegetation and substrate in the present study. Vegetation, transparency, depth, and substratum covary naturally in lakes; therefore, their effects on the distribution of yellow perch may have been confounded in previous studies that did not consider covariates, latent spatial effects, and temporal effects simultaneously.

The model that performed best (M9) incorporated both anisotropy and a separate Gaussian process for each shore. The model comparisons point to marked differences in the posterior structure of latent spatial effects for the two shores: anisotropy was conspicuous in the north shore, whereas spatial structuring was weak in the south shore. The rate of decay in spatial correlation in the north shore had marked directionality and was generally slower in the SW - NE direction, broadly in alignment with the shoreline (Fig. \ref{fig:gam}b). Likely environmental candidates responsible for these intershore differences in latent spatial effects include known differences in physical and chemical properties of water masses between the two shores, which are influenced by contributions from different tributaries and diffuse sources of nutrients and pollutants. Exposure to more differentiated water masses in the north shore has been invoked as an explanation for the greater variability in growth of yellow perch on the north shore of the lake \citep{glemet:rodriguez:2007}. The importance of local effects indicates that traditional covariates such as water depth and transparency provide only a partial picture of environmental influences on the spatial distribution of yellow perch in the lake. Further research on the effects of shoreline-dependent processes and larval transport from the spawning grounds may therefore prove fruitful.

Although fish sampling was carefully standardized, we cannot rule out the possibility that sampling efficiency and thus, probability of detection, were influenced by environmental characteristics, which could confound inference on environmental effects. Identifying environmental effects on abundance separately from those on detection in an open population of unmarked individuals would require more complex sampling designs and observational models, such as $N$-mixture models \citep{royle:2004}. However, these models invoke an assumption of closure over repeated observations that may be difficult to justify when using active gear to sample populations of highly mobile fish.

The approaches discussed here might pose computational challenges when the number of locations is large. In such cases, alternative approaches based on Gaussian random Markov fields \citep{Lindgren2011} may be useful to provide a prior distribution for $Z$.

An important feature of the present study is its treatment of non-standard spatial features of the example, including the spatial arrangement of samples along the lake shoreline and the presence of the channel running through the lake. The circular models are expected to be most applicable in situations in which the spatial domain is approximately oval, but more flexible anisotropic models, such as M10, may perform better under more complex domain topologies. Concerns about the influence of irregular domain shapes and complex boundaries on the outcome of spatial analyses have been voiced previously (\citealp{legates:1991}; \citealp{ramsay2002}; \citealp{soubeyrand:enjalbert:sache:2008}; \citealp{miller:2014}). Recently developed methods, such as the soap film smoother (\citealp{wood2008}) and generalized distance splines (\citealp{miller:2014}), are promising alternative approaches for dealing effectively with complex irregular boundaries or interior holes. However, little effort has been devoted to developing approaches that systematically compare competing representations of the spatial domain. The approaches presented here for specification of spatial domain and choice of Gaussian process priors may prove useful in other applications that involve spatial correlation along regular, possibly discontinous, contours. Biological examples include samples collected along a mountain perimeter within an altitudinal range, at different heights on the bark of a tree trunk, along the edges of growing structures (e.g., bacterial colonies, diffusing chemicals), and at interfaces between habitats (ecotones).

\section*{Acknowledgements}
We thank the anonymous reviewers, Tilmann Gneiting, and the associate editor for their constructive comments, and  Ephraim Hanks for helpful suggestions on the implementation of restricted spatial regression. We are grateful for financial support from CNPq and FAPERJ (A.M. Schmidt), the Natural Sciences and Engineering Council of Canada, CAPES and FAPERJ (M.A. Rodr\'iguez), and CAPES (scholarship to E.S. Capistrano).

\begin{supplement}[id=suppA]
%\sname{}
\stitle{Additional results for ``Population counts along elliptical habitat contours: hierarchical modelling using Poisson-lognormal mixtures with nonstationary spatial structure"} \slink[doi]{COMPLETED BY THE TYPESETTER} \sdatatype{.pdf} \sdescription{This supplement contains four sections which provide further results on: 1) circular transformations, 2) model comparison criteria, 3) analyses of model fit and correlation of local effects, and 4) restricted spatial regression.}
\end{supplement}

\bibliographystyle{newapa}
\bibliography{librox}

\begin{thebibliography}{}

\bibitem[\protect\citeauthoryear{Bertolo, Blanchet, Magnan, Brodeur, Mingelbier
  \& Legendre}{Bertolo et~al.}{2012}]{bertolo:2012}
Bertolo, A., Blanchet, A. F.~G., Magnan, P., Brodeur, P., Mingelbier, M., \&
  Legendre, P. (2012).
\newblock Inferring processes from spatial patterns: the role of directional
  and non-directional forces in shaping fish larvae distribution in a
  freshwater lake system.
\newblock {\em PLoS ONE}, {\em 7}, 1--11.

\bibitem[\protect\citeauthoryear{Bulmer}{Bulmer}{1974}]{bulmer:1974}
Bulmer, M.~G. (1974).
\newblock On fitting the {P}oisson lognormal distribution to species-abundance
  data.
\newblock {\em Biometrics}, {\em 30}, 101--110.

\bibitem[\protect\citeauthoryear{Clark \& Gelfand}{Clark \&
  Gelfand}{2006}]{clark:gelfand:2006}
Clark, J.~S. \& Gelfand, A.~E. (2006).
\newblock {\em Hierarchical Modelling for the Environmental Sciences:
  Statistical Methods and Applications}.
\newblock Oxford University Press, Oxford, UK.

\bibitem[\protect\citeauthoryear{Czado, Gneiting \& Held}{Czado
  et~al.}{2009}]{czado:gneiting:held:2009}
Czado, C., Gneiting, T., \& Held, L. (2009).
\newblock Predictive model assessment for count data.
\newblock {\em Biometrics}, {\em 65}, 1254--1261.

\bibitem[\protect\citeauthoryear{Denison \& Mallick}{Denison \&
  Mallick}{1998}]{denison:mallick:1998}
Denison, D. G.~T. \& Mallick, B.~K. (1998).
\newblock Discussion of model-based geostatistics.
\newblock {\em Applied Statistics}, {\em 47}, 336.

\bibitem[\protect\citeauthoryear{Diggle, Moyeed \& Tawn}{Diggle
  et~al.}{1998}]{diggle:moyeed:tawn:1998}
Diggle, P., Moyeed, R., \& Tawn, J. (1998).
\newblock Model-based geostatistics (with discussion).
\newblock {\em Applied Statistics}, {\em 47}, 299--350.

\bibitem[\protect\citeauthoryear{Diggle \& Ribeiro}{Diggle \&
  Ribeiro}{2007}]{diggle:ribeiro:2007}
Diggle, P.~J. \& Ribeiro, P.~J. (2007).
\newblock {\em Model-Based Geostatistics}.
\newblock Springer, New York, USA.

\bibitem[\protect\citeauthoryear{Doornik}{Doornik}{2007}]{doornik:2007}
Doornik, J. (2007).
\newblock {\em Object-Oriented Matrix Programming Using Ox, 3rd edition}.
\newblock Timberlake Consultants Press and Oxford, London, UK.

\bibitem[\protect\citeauthoryear{Gamerman}{Gamerman}{1997}]{gamerman:1997}
Gamerman, D. (1997).
\newblock Sampling from the posterior distribution in generalized linear mixed
  models.
\newblock {\em Statistics and Computing}, {\em 7}, 57--68.

\bibitem[\protect\citeauthoryear{Gl\'emet \& Rodr\'{\i}guez}{Gl\'emet \&
  Rodr\'{\i}guez}{2007}]{glemet:rodriguez:2007}
Gl\'emet, H. \& Rodr\'{\i}guez, M.~A. (2007).
\newblock Short-term growth ({RNA}/{DNA} ratio) of yellow perch
  (\textit{{P}erca flavescens}) in relation to environmental influence and
  spatio-temporal variation in a shallow fluvial lake.
\newblock {\em Canadian Journal of Fisheries and Aquatic Sciences}, {\em 64},
  1646--1655.

\bibitem[\protect\citeauthoryear{Gneiting}{Gneiting}{2013}]{gneiting:2013}
Gneiting, T. (2013).
\newblock Strictly and non-strictly positive definite functions on spheres.
\newblock {\em Bernoulli}, {\em 19}, 1327--1349.

\bibitem[\protect\citeauthoryear{Gneiting \& Raftery}{Gneiting \&
  Raftery}{2007}]{gneiting:raftery:2007}
Gneiting, T. \& Raftery, A.~E. (2007).
\newblock Strictly proper scoring rules, prediction and estimation.
\newblock {\em Journal of the American Statistical Association}, {\em 102},
  359--378.

\bibitem[\protect\citeauthoryear{Gschl\"{o}{\ss}l \& Czado}{Gschl\"{o}{\ss}l \&
  Czado}{2008}]{gsch:czad:2008}
Gschl\"{o}{\ss}l, S. \& Czado, C. (2008).
\newblock Modelling count data with overdispersion and spatial effects.
\newblock {\em Statistical Papers}, {\em 49}, 531--552.

\bibitem[\protect\citeauthoryear{Guttorp \& Schmidt}{Guttorp \&
  Schmidt}{2013}]{guttorp:schmidt:2013}
Guttorp, P. \& Schmidt, A.~M. (2013).
\newblock Covariance structure of spatial and spatio-temporal processes.
\newblock {\em WIREs Computational Statistics}, {\em 5}, 279--287.

\bibitem[\protect\citeauthoryear{Hanks, Schliep, Hooten \& Hoeting}{Hanks
  et~al.}{2015}]{Hanks:Schliep:Hooten:Hoeting:2014}
Hanks, E.~M., Schliep, E.~M., Hooten, M.~B., \& Hoeting, J.~A. (2015).
\newblock Restricted spatial regression in practice: geostatistical models,
  confounding, and robustness under model misspecification.
\newblock {\em Environmetrics}, {\em 26}, 243--254.

\bibitem[\protect\citeauthoryear{Hastings}{Hastings}{1970}]{hastings:1970}
Hastings, W.~K. (1970).
\newblock Monte {C}arlo sampling methods using {M}arkov chains and their
  applications.
\newblock {\em Biometrika}, {\em 57}, 97--109.

\bibitem[\protect\citeauthoryear{Hodges \& Reich}{Hodges \&
  Reich}{2010}]{Hodges:Reich:2010}
Hodges, J.~S. \& Reich, B.~J. (2010).
\newblock Adding spatially-correlated errors can mess up the fixed effect you
  love.
\newblock {\em The American Statistician}, {\em 64}, 325--334.

\bibitem[\protect\citeauthoryear{Hudon}{Hudon}{1997}]{hudon:1997}
Hudon, C. (1997).
\newblock Impact of water level fluctuations on {S}t. {L}awrence {R}iver
  aquatic vegetation.
\newblock {\em Canadian Journal of Fisheries and Aquatic Sciences}, {\em 54},
  2853--2865.

\bibitem[\protect\citeauthoryear{Hughes \& Haran}{Hughes \&
  Haran}{2013}]{Hughes:Haran:2013}
Hughes, J. \& Haran, M. (2013).
\newblock Dimension reduction and alleviation of confounding for spatial
  generalized linear mixed models.
\newblock {\em Journal of the Royal Statistical Society, Series B}, {\em 75},
  139--159.

\bibitem[\protect\citeauthoryear{Ingebrigtsen, Lindgren \&
  Steinsland}{Ingebrigtsen
  et~al.}{2014}]{ingebrigtsen:lindgren:steinsland:2014}
Ingebrigtsen, R., Lindgren, F., \& Steinsland, I. (2014).
\newblock Spatial models with explanatory variables in the dependence
  structure.
\newblock {\em Spatial Statistics}, {\em 8}, 20--38.

\bibitem[\protect\citeauthoryear{Legates}{Legates}{1991}]{legates:1991}
Legates, D.~R. (1991).
\newblock The effect of domain shape on principal components analyses.
\newblock {\em International Journal of Climatology}, {\em 11}, 135--146.

\bibitem[\protect\citeauthoryear{Lindgren, Rue \& Lindstr\"{o}m}{Lindgren
  et~al.}{2011}]{Lindgren2011}
Lindgren, F., Rue, H., \& Lindstr\"{o}m, J. (2011).
\newblock An explicit link between {G}aussian fields and {G}aussian {M}arkov
  random fields: the stochastic partial differential equation approach.
\newblock {\em Journal of the Royal Statistical Society, Series B}, {\em 73},
  423--498.

\bibitem[\protect\citeauthoryear{Metropolis, Rosenbluth, Rosenbluth, Teller \&
  Teller}{Metropolis
  et~al.}{1953}]{metropolis:rosenbluth:rosenbluth:teller:teller:1953}
Metropolis, N., Rosenbluth, A.~W., Rosenbluth, M.~N., Teller, A.~H., \& Teller,
  E. (1953).
\newblock Equation of state calculations by fast computing machines.
\newblock {\em Journal of Chemical Physics}, {\em 21}, 1087--1092.

\bibitem[\protect\citeauthoryear{Miller \& Wood}{Miller \&
  Wood}{2014}]{miller:2014}
Miller, D.~L. \& Wood, S. (2014).
\newblock Finite area smoothing with generalized distance splines.
\newblock {\em Environmental and Ecological Statistics}, {\em 21}, 1--17.

\bibitem[\protect\citeauthoryear{Paciorek}{Paciorek}{2010}]{Paciorek:2010}
Paciorek, C.~J. (2010).
\newblock The importance of scale for spatial-confounding bias and precision of
  spatial regression estimators.
\newblock {\em Statistical Science}, {\em 25}, 107--125.

\bibitem[\protect\citeauthoryear{Plummer, Best, Cowles \& Vines}{Plummer
  et~al.}{2006}]{plummer:2006}
Plummer, M., Best, N., Cowles, K., \& Vines, K. (2006).
\newblock {CODA}: Convergence diagnosis and output analysis for {MCMC}.
\newblock {\em R News}, {\em 6}, 7--11.

\bibitem[\protect\citeauthoryear{Poppick \& Stein}{Poppick \&
  Stein}{2014}]{poppick:stein:2014}
Poppick, A. \& Stein, M.~L. (2014).
\newblock Using covariates to model dependence in nonstationary, high-frequency
  meteorological processes.
\newblock {\em Environmetrics}, {\em 25}, 293--305.

\bibitem[\protect\citeauthoryear{Ramsay}{Ramsay}{2002}]{ramsay2002}
Ramsay, T. (2002).
\newblock Spline smoothing over difficult regions.
\newblock {\em Journal of the Royal Statistical Society: Series B (Statistical
  Methodology)}, {\em 64}, 307--319.

\bibitem[\protect\citeauthoryear{Reich, Hodges \& Zadnik}{Reich
  et~al.}{2006}]{Reich:Hodges:Zadnik:2006}
Reich, B.~J., Hodges, J.~S., \& Zadnik, V. (2006).
\newblock Effects of residual smoothing on the posterior of the fixed effects
  in disease-mapping models.
\newblock {\em Biometrics}, {\em 62}, 1197--1206.

\bibitem[\protect\citeauthoryear{Roberts \& Rosenthal}{Roberts \&
  Rosenthal}{2009}]{roberts:rosenthal:2009}
Roberts, G.~O. \& Rosenthal, J.~S. (2009).
\newblock Examples of adaptive {MCMC}.
\newblock {\em Journal of Computational and Graphical Statistics}, {\em 18},
  349--367.

\bibitem[\protect\citeauthoryear{Royle}{Royle}{2004}]{royle:2004}
Royle, J.~A. (2004).
\newblock N-mixture models for estimating population size from spatially
  replicated counts.
\newblock {\em Biometrics}, {\em 60}, 108--115.

\bibitem[\protect\citeauthoryear{Sampson \& Guttorp}{Sampson \&
  Guttorp}{1992}]{samp:gut:1992}
Sampson, P. \& Guttorp, P. (1992).
\newblock Nonparametric estimation of nonstationary spatial covariance
  structure.
\newblock {\em Journal of the American Statistical Association}, {\em 87},
  108--119.

\bibitem[\protect\citeauthoryear{Schmidt, Guttorp \& O'Hagan}{Schmidt
  et~al.}{2011}]{schmidt:guttorp:ohagan:2011}
Schmidt, A.~M., Guttorp, P., \& O'Hagan, A. (2011).
\newblock Considering covariates in the covariance structure of spatial
  processes.
\newblock {\em Environmetrics}, {\em 22}, 487--500.

\bibitem[\protect\citeauthoryear{Schmidt \& Rodr\'iguez}{Schmidt \&
  Rodr\'iguez}{2011a}]{schmidt:rodriguez:2011a}
Schmidt, A.~M. \& Rodr\'iguez, M.~A. (2011a).
\newblock Modelling multivariate counts varying continuously in space (with
  discussion).
\newblock In Bernardo, J.~M., Bayarri, M.~J., Berger, J.~O., Dawid, A.~P.,
  Heckerman, D., Smith, A. F.~M., \& West, M. (Eds.), {\em Bayesian Statistics
  9}, (pp.\ 611--638). Oxford University Press, Oxford, UK.

\bibitem[\protect\citeauthoryear{Schmidt \& Rodr\'iguez}{Schmidt \&
  Rodr\'iguez}{2011b}]{schmidt:rodriguez:2011b}
Schmidt, A.~M. \& Rodr\'iguez, M.~A. (2011b).
\newblock Reply to the discussion of {B}oys, {F}arrow, and {G}ermain.
\newblock In Bernardo, J.~M., Bayarri, M.~J., Berger, J.~O., Dawid, A.~P.,
  Heckerman, D., Smith, A. F.~M., \& West, M. (Eds.), {\em Bayesian Statistics
  9}, (pp.\ 630--638). Oxford University Press, Oxford, UK.

\bibitem[\protect\citeauthoryear{Schmidt, Rodr\'iguez \& Capistrano}{Schmidt
  et~al.}{2015}]{schmidt:rodriguez:capistrano}
Schmidt, A.~M., Rodr\'iguez, M.~A., \& Capistrano, E.~S. (2015).
\newblock Supplement to {P}opulation counts along elliptical habitat contours:
  hierarchical modelling using {P}oisson-lognormal mixtures with nonstationary
  spatial structure.
\newblock {\em Annals of Applied Statistics}.

\bibitem[\protect\citeauthoryear{Scott \& Crossman}{Scott \&
  Crossman}{1973}]{scott:crossman:1973}
Scott, W.~B. \& Crossman, E.~J. (1973).
\newblock {\em Freshwater Fishes of Canada}.
\newblock Fisheries Research Board of Canada, Bulletin 184, Ottawa, Canada.

\bibitem[\protect\citeauthoryear{Soubeyrand, Enjalbert \& Sache}{Soubeyrand
  et~al.}{2008}]{soubeyrand:enjalbert:sache:2008}
Soubeyrand, S., Enjalbert, J., \& Sache, I. (2008).
\newblock Accounting for roughness of circular processes: using {G}aussian
  random processes to model anisotropic spread of airborne plant disease.
\newblock {\em Theoretical Population Biology}, {\em 73}, 92--103.

\bibitem[\protect\citeauthoryear{Spiegelhalter, Best, Carlin \&
  Linde}{Spiegelhalter et~al.}{2002}]{spiegelhalter:best:carlin:linde:2002}
Spiegelhalter, D., Best, N., Carlin, B., \& Linde, A. (2002).
\newblock {B}ayesian measures of model complexity and fit (with discussion).
\newblock {\em Journal of the Royal Statistical Society. {B}}, {\em 64},
  583--639.

\bibitem[\protect\citeauthoryear{Wikle}{Wikle}{2003}]{wikle:2003}
Wikle, C.~K. (2003).
\newblock Hierarchical models in environmental science.
\newblock {\em International Statistical Review}, {\em 71}, 181--199.

\bibitem[\protect\citeauthoryear{Wikle}{Wikle}{2010}]{wikle:2010}
Wikle, C.~K. (2010).
\newblock Hierarchical modeling with spatial data.
\newblock In A.~Gelfand, P.~Diggle, M.~Fuentes, \& P.~Guttorp (Eds.), {\em
  Handbook of Spatial Statistics}  (pp.\ 89--105). Chapman \& Hall/CRC.

\bibitem[\protect\citeauthoryear{Williams}{Williams}{1998}]{williams:1998}
Williams, C. K.~I. (1998).
\newblock Discussion of model-based geostatistics.
\newblock {\em Applied Statistics}, {\em 47}, 342.

\bibitem[\protect\citeauthoryear{Wood}{Wood}{2006}]{wood:2006}
Wood, S.~N. (2006).
\newblock {\em Generalized Additive Models: An Introduction with R}.
\newblock Chapman \& Hall, New York.

\bibitem[\protect\citeauthoryear{Wood, Bravington \& Hedley}{Wood
  et~al.}{2008}]{wood2008}
Wood, S.~N., Bravington, M.~V., \& Hedley, S.~L. (2008).
\newblock Soap film smoothing.
\newblock {\em Journal of the Royal Statistical Society: Series B (Statistical
  Methodology)}, {\em 70}, 931--955.

\bibitem[\protect\citeauthoryear{Yaglom}{Yaglom}{1987}]{yaglom:1987}
Yaglom, A.~M. (1987).
\newblock {\em Correlation Theory of Stationary and Related Random Functions I
  - Basic Results}.
\newblock Springer-Verlag, New York.

\end{thebibliography}

\end{document}